\documentclass[12pt,epsfig]{article}
\usepackage[english]{babel}
\usepackage{epsfig}

\begin{document}

\title{\small COULOMB SINGULARITY EFFECTS IN TUNNELLING SPECTROSCOPY OF INDIVIDUAL IMPURITIES}
\author{\small P.I. Arseyev, N.S. Maslova, V.I. Panov, S.V. Savinov. \\
\small Department of Physics, Moscow State University, 119899 Moscow, Russia}
\date{}
\maketitle


\begin{abstract}
Non-equilibrium Coulomb effects in resonant tunnelling through deep impurity
states are analyzed. It is shown that  Coulomb
vertex corrections to the tunnelling transfer
amplitude
lead to power law singularity in current-voltage
characteristics.
\end{abstract}

Localized states of individual impurity atoms  and interacting impurity
clusters can play the key role in tunnelling processes in small size
junctions  and often determine the behavior of tunnelling characteristics
in STM/STS contacts. Now it is evident that in tunnelling junctions of
nanometer scale there exists non-equilibrium distribution of tunnelling
electrons which changes  local
density of states and tunnelling conductivity spectra. Some interesting
effects, such as resonance structure of tunnelling conductivity inside
semiconductor band gap, increased value of observed band gap and
non-equilibrium interaction of neighboring impurity atoms have been
recently investigated experimentally and theoretically analyzed
~\cite{1,2,3,4}.  But all these effects are caused by local changes of the
initial
density of states in the contact area due to
interactions of non-equilibrium particles. The modification of tunnelling
amplitude by the Coulomb interaction of conduction electrons in metallic tip
with non-equilibrium localized charges was ignored. It is shown in
the present paper that corrections to the
tunnelling vertex caused by the Coulomb potential
can also
result in nontrivial behavior of tunnelling characteristics and should be
taken into account.  One  encounters  with effects similar to the Mahan
edge singularities in the problem of X - ray absorption spectra in metals ~\cite{5}.
The effect is well pronounced if
tunnelling rate from a deep impurity level to metallic tip $\gamma_t$ is
much larger than  relaxation rate $\gamma$ of non-equilibrium electron
distribution at localized state. This condition can be realized experimentally
for a deep impurity state in the semiconductor gap.
Direct tunnelling from such states to semiconductor continuum states  is
strongly reduced due to wide barrier formed by surface band bending.
Relaxation rate connected with electron-phonon interaction
can be estimated to be of the order  $10^8 - 10^{10} 1/s$
at low temperatures~\cite{5.5}.
As to $\gamma_t$ --- it is parameter, which can be varied in STM/STS experiments
changing tip-sample separation. Since tip-sample separation is comparable with
atomic scale,  $\gamma_t$ often exceeds the relaxation rate for
deep impurity states.
Typical experimental value of
tunnelling current 1 nA corresponds to $\gamma_t \simeq 10^{11} - 10^{12} 1/s$
 ~\cite{3}.
As it will be shown below for $\gamma_t \gg \gamma $ the impurity
 level becomes nearly empty when the value of applied bias voltage
 approaches the impurity energy.
So the core hole Coulomb potential is suddenly switched on and the
tunnelling amplitude is changed.
One might expect in this situation a
power-law singularity in current-voltage characteristics
near the threshold voltage.

The system semiconductor - impurity state - metallic tip can be described
by the Hamiltonian
 $\hat{H}$:
\begin{equation}
 \hat{H} =  \hat{H}_L +  \hat{H}_R +  \hat{H}_{imp} +  \hat{H}_T +  \hat{H}_{int}
\end{equation}
where:
\begin{equation}
 \hat{H}_R = \sum_{k\sigma} { (\varepsilon_k -\mu ) c_{k \sigma}^+ c_{k \sigma}
 },
 \hat{H}_L = \sum_{p\sigma} { (\varepsilon_p -\mu - eV) c_{p \sigma}^+ c_{p \sigma} }
\end{equation}
describes the electron states in the metallic tip and the semiconductor
correspondingly,
  $c_{k \sigma}^+(c_{k \sigma})$ and $c_{p \sigma}^+(c_{p \sigma})$ describe creation (annihilation) of electron in
states $(k\sigma)$ and $(p\sigma)$ in each bank of the contact.
\begin{equation}
 \hat{H}_{imp} = \sum_{d\sigma} { \varepsilon_d  c_{d \sigma}^+ c_{d \sigma} + U n_{d \sigma}n_{d {-\sigma}}}
\end{equation}
corresponds to a localized impurity state.  We consider "one electron
neutral impurity" - the impurity level is single occupied at zero applied
voltage due to the on-site Coulomb interaction.  But
in the case of large tunnelling rate to
the metallic tip  the  on-site Coulomb repulsion of
localized electrons can be omitted,
if we analyze the
behavior of tunnelling current at applied voltage close to the impurity
energy $\varepsilon_d$. Because in this situation the impurity state
becomes nearly
empty above the threshold value of the applied bias.  Let us also point out,
that the Kondo regime is
destroyed  on Anderson impurity
 for the values of applied bias near the threshold ~\cite{8,9}.
 In this case the Kondo-effect is not responsible for any unusual
features of the tunnelling characteristics .

Tunnelling transitions from the impurity state to the semiconductor and
the metal are described by the part:
\begin{equation}
 \hat{H}_T = \sum_{k  p} {(T_{kd} c_{k \sigma}^+ c_{d \sigma}  + T_{pd} c_{p \sigma}^+ c_{d
 \sigma})}+ h. c.
\end{equation}
And, finally, part $H_{int}$ includes the Coulomb interaction of the core
(impurity) hole with conduction electrons in the metal.
\begin{equation}
 \hat{H}_{int} = \sum_{k k' \sigma \sigma'} {W_{kk'}
  c_{k \sigma}^+ c_{k' \sigma}(1-c_{d\sigma'}^+ c_{d\sigma'}) }
\end{equation}
Hamiltonian $H_{int}$ appears as many particle interaction and describes
rearrangement of conduction electrons in the potential of the hole,
suddenly switched on by tunnelling transition of the impurity electron. Since we
are far from the Kondo regime it is sufficient to consider tunnelling current
in the lowest order in the tunnelling amplitude $T_{kd}$.  Scattering by
the impurity hole Coulomb potential does not change electron spin.  Thus
in
the lowest order in $T_{kd}$ we can consider renormalization
of the tunnelling amplitude
independently for each spin - the same one for conduction and impurity
electrons.   It is also reasonable to use for simlicity an
averaged value of screened
Coulomb interaction describing s - wave scattering of conduction
 electrons by a deep hole $W_{kk'} = W$.

 Edge singularities in the tunnelling current can be analyzed by means
 of diagram technique for non-equilibrium
 processes.
Using Keldysh functions $G^<$ the tunnelling current can be determined as
(we set charge $e=1$):
\begin{equation}
\label{I}
I(V)=Im(J(V))\qquad , \qquad
  J(V) = i\sum_{k,  \sigma} \int{ d\omega} T_{kd} G^{\sigma<}_{kd}
\end{equation}
where we have defined tunnelling "response function" $J(V)$.
If the Coulomb interaction is neglected one can obtain the usual expression for this
response function in the lowest order in $T_{kd}$:
\begin{equation}
    \label{J0G}
  J^0(V) = i\sum_{k,  \sigma} \int{ d\omega} T_{kd}^2( G^{\sigma<}_{kk} G^{\sigma A}_{dd}+ G^{\sigma R}_{kk} G^{\sigma<}_{dd})
\end{equation}
Substituting the correspondent expressions for the Keldysh functions
 ~\cite{4}  and performing integration over $k$ we get:
\begin{equation}
  \label{J0}
  J^0(V) = \gamma_t \int{ d\omega}\left[
  \frac {n_k^0(\omega)}
  {\omega + eV - \varepsilon_d + i(\gamma + \gamma_t)}
+
   \frac {n_d(\omega)(-i(\gamma + \gamma_t))}
  {(\omega + eV - \varepsilon_d)^2 + (\gamma + \gamma_t)^2}\right]
\end{equation}
where the tunnelling rate $\gamma_t = T_{kd}^2\nu  $, and $\nu$ is unperturbed
density of states in the metallic tip. Kinetic parameter $\gamma$ corresponds
to relaxation rate of electron distribution at the localized state.
In the suggested microscopic picture (eq.(4)) this
relaxation rate
is
determined by  small enough electron tunnelling transitions from the impurity
to the semiconductor continuum states
$\gamma = T_{pd}^2\nu_p$. (In general $\gamma$ can include different types
of relaxation processes.)

Non-equilibrium impurity filling numbers $n_d(\omega)$ are determined from kinetic
equations for the Keldysh functions $G^<$:
\begin{equation}
n_d(\omega)= \frac{\gamma n_p^0(\omega)+\gamma_t n_k^0(\omega)  }{\gamma +\gamma_t}
\end{equation}
As it was explained in the introduction
for a deep impurity level the relation $\gamma_t >> \gamma$ is quite possible.
Then $n_p^0(\varepsilon_d)
= 1 $, while $n_d(\varepsilon_d) << 1$ and there is really a core hole in
the impurity state.
 Thus for low temperatures one can obtain from Eq.(\ref{J0}):
\begin{equation}
     \label{J0ln}
  J^0(V) = \gamma_t \, ln(|X|)+ i \frac{ \gamma_t\gamma  }{\gamma + \gamma_t}
  \left[ arcctg((eV - \varepsilon_d)/(\gamma +
\gamma_t))- arcctg((- \varepsilon_d)/(\gamma +
\gamma_t))\right]
\end{equation}
where
\begin{equation}
 X = ( eV - \varepsilon_d + i(\gamma + \gamma_t))/D)
\end{equation}
and $D$ is the band width for electrons in metal.

The usual form of the tunnelling current
is of course reproduced from  Eqs.(\ref{I},\ref{J0},\ref{J0ln}).

Now let us consider renormalization of the tunnelling amplitude and vertex
corrections to the tunnelling current caused by the Coulomb
interaction between the impurity core hole and electrons in the metal.
Many particle
picture strongly differs from the single-particle one near the threshold
voltage.  First order corrections
due to the Coulomb interaction (the first graph in Fig. 1(a)) has logarithmic
divergency at the threshold voltage $eV = \varepsilon_d $, which is cut off
by the finite relaxation and tunnelling rates.

\begin{equation}
    \label{J1}
  J^1(V) = i\sum_{k,  \sigma} \int{ d\omega} T_{kd}(- G^{\sigma<}_{kk} G^{\sigma A}_{dd}T_{kd}^{1++}+ G^{\sigma R}_{kk} G^{\sigma<}_{dd}T_{kd}^{1--})
\end{equation}

Tunnelling matrix elements are changed by the Coulomb interaction:
\begin{equation}
  \label{T1}
  T_{kd}^{1 --} = \sum_{k,  \sigma} \int{ d\omega} T_{kd}W
    ( G^{\sigma<}_{kk} G^{\sigma A}_{dd}+ G^{\sigma R}_{kk} G^{\sigma<}_{dd})
\end{equation}

If we look at Eqs. (\ref{J0G},\ref{J0}),  it becomes clear, that
logarithmic contribution comes from the first combination of the Green functions:
$ G^{\sigma<}_{kk} G^{\sigma A}_{dd}$.
In what follows we retain only logarithmically large parts, assuming that
$|ln((\gamma + \gamma_t))/D)| \gg 1$,
so only these combinations of the Green functions are the most important
in perturbation series.
Then from (\ref{T1}) we obtain that
tunnelling amplitude contains logarithmic correction: $ T_{kd}^{1 --} = -T_{kd} L$,
$ T_{kd}^{1 ++} = -T_{kd} ^{1--}$,
where factor $L$:

\begin{equation}
L =  (W \nu) ln(X)
\end{equation}
\begin{figure}
\leavevmode
\centering{\psfig{figure=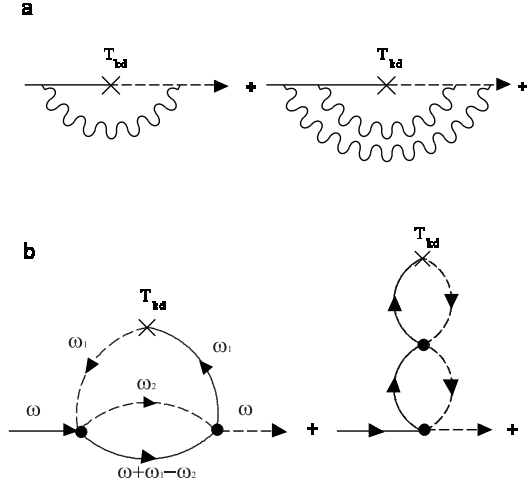,height=10cm}}
\caption{Coulomb corrections to $T_{kd}$. Solid lines represent $G_k$ and dashed
lines --- $G_d$.
 a) Ladder approximation,
 b) Parquet graphs (Coulomb wavy lines are overdrawn as black circle vertexes)
  }
\label{Fig.1}
\end{figure}

In high orders of perturbation expansion ladder graphs (Fig.1 a)
are the simplest "maximally singular" graphs.
But this is not the only relevant kind of graphs.
If we look at the first graph in Fig.1b, we notice, that
a new type of "bubble" appears, which is logarithmically large
for small "total" energy $(\omega + \omega_1)$.
The important point is that relevant region of
integration  over $\omega$ and $\omega_1$
is region of small $\omega$.  It is just this region which gives essential
contribution to logarithmic factor $L$ in any other pair of
$ G^{<}_{kk} G^{A(R)}_{dd}$.

It means that the central bubble also
contributes an additional logarithmic factor to the total result.
In this situation, which is not new in physics, one should retain in
the n - th order of perturbation expansion
the most divergent terms
proportional to $ (W\nu)^n L^{n+1}$ .
For the first time such method was developed by Dyatlov et al ~\cite{5.6}.
It was shown that
for a proper
treatment of this problem one should write down integral equations for the
so-called
parquet graphs (Fig.  1b), which are constructed by successive
substitution the simple Coulomb vertex  for the two types
of bubbles in perturbation
series.
These equations represent some extension of the ordinary
Bethe-Salpeter equation and
describe multiple scattering of conduction
electrons by the core hole Coulomb potential in  the two "most singular"
channels. The integral equations can be solved with
logarithmic accuracy, as it was done,
for example, by Nozieres ~\cite{6,7} for edge singularities in X - ray absorption
spectra in metals.

The solution of the "parquet" equations contains nothing new in our problem.
So we present the result without getting into technical details.
Summing up the most divergent graphs with logarithmic accuracy
one can obtain the following singular part of the
response function ~\cite{6}:

\begin{equation}
J(V)  =  \frac{ \gamma_t( 1 - exp (- 2L))}{2W \nu}
\end{equation}

Then the tunnelling current near the threshold voltage can be expressed as:

\begin{equation}
  \label{If}
  I(V) = \frac{ \gamma_t}{2W \nu  }
  \left[\frac {D^2}
{(eV - \varepsilon_d)^2+ (\gamma_t +\gamma)^2 }\right]
^{W \nu } sin\left(2 W \nu \phi \right)
\end{equation}

where $ \phi = arcctg (\frac{eV -\varepsilon_d }{\gamma + \gamma_t})$.
If we consider a deep impurity state in the gap of the
semiconductor (below the Fermi level)
and
positive tip bias voltage, then: $\varepsilon_d < 0, eV < 0$.  So, the
phase $\phi$ is a step-like function
varying approximatly from $0$ to $\pi$, when the applied bias crosses
the threshold $eV =\varepsilon_d $.
 Since we retain only the
most logarithmically  large terms in the tunnelling current,  Eq.(\ref{If})
is valid only if $ |eV -\varepsilon_d| \ll D$.
In the absence of the Coulomb interaction ($W=0$)
this singular part reduces to the usual first order contribution
arising from the first term in Eqs.(\ref{J0G},\ref{J0}).

 In summary, if
the tunnelling rate to the metallic tip exceeds
the relaxation rate of localized  electrons,
the Coulomb interaction of the core hole and the conduction electrons in the
metal
strongly modifies the tunnelling transition amplitude and leads to:
i)non-monotonic behavior of current voltage characteristics; ii)power-law
singularity of the tunnelling current and conductivity when the value of
the applied voltage approaches the impurity level energy; iii) current voltage
characteristics can be rather asymmetric, because of different dependence
of the phase factor $\phi$ on the applied bias below and above the threshold
value.
Power law singular behavior of the tunnelling current is sensitive
to the values of the tunnelling and relaxation rates, as well as to the value
of the Coulomb interaction $W$.
So, different exponents in power dependencies
of the tunnelling current on the applied voltage can appear with changing tip-
sample separation. Some current-voltage characteristics obtained for typical
values of parameters  are shown
in Fig. 2.

\begin{figure}
\leavevmode
\centering{\psfig{figure=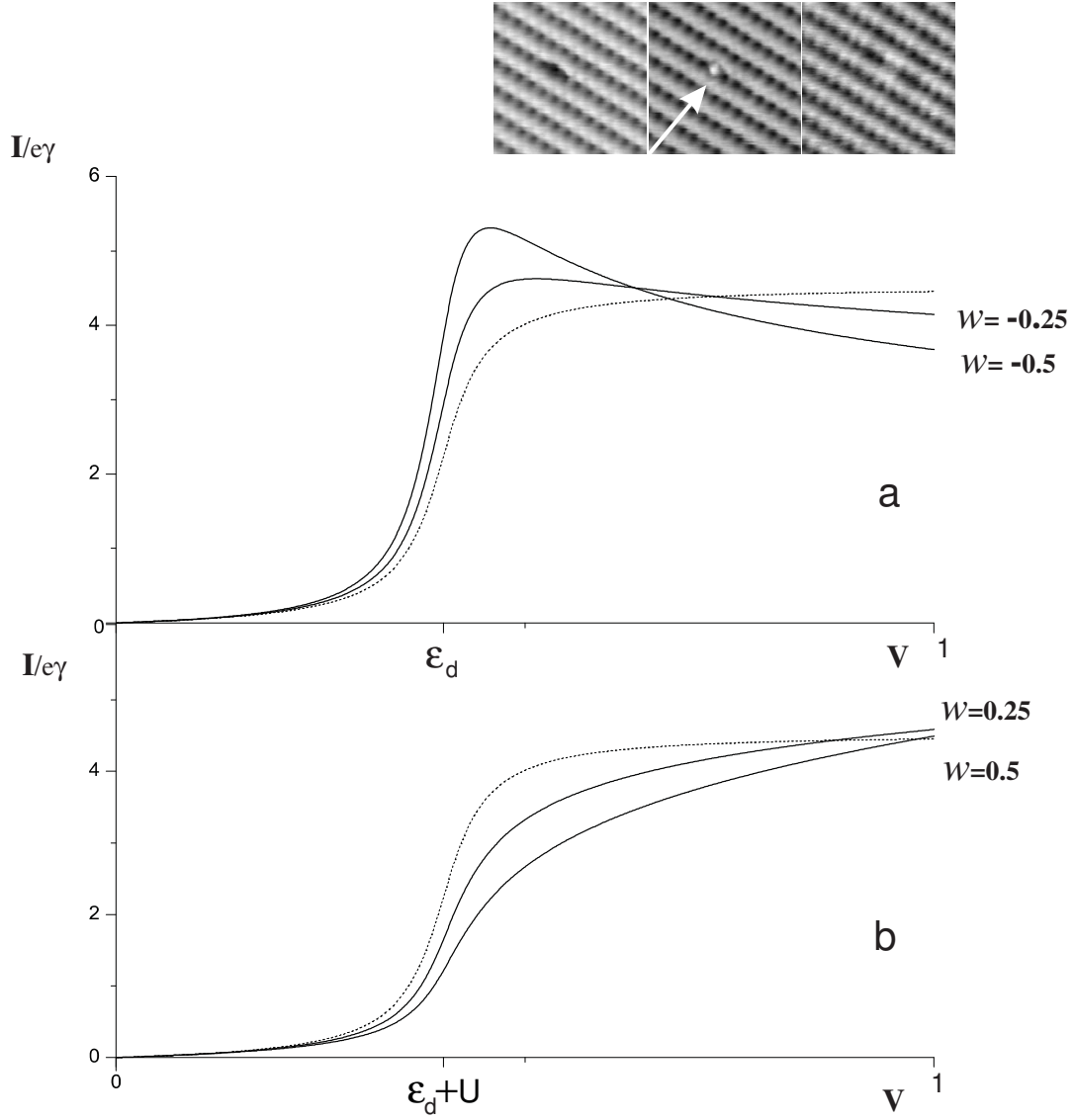,height=15cm}}
\caption{Current-voltage curves for typical values of dimensionless Coulomb and kinetic
parameters. Current is measured in dimensionless units $I/e\gamma$.
a) $w=W \nu < 0, \varepsilon_d=0.4eV$, b) $w=W \nu > 0 , \varepsilon_d+U=0.4eV$,
$\gamma_t/\gamma =3, \varepsilon_d/\gamma =40$. Dashed lines correspond to
$W=0$. Experimental STM image of Cr impurity on InAs (110) surface
 is shown in the inset for
 V=0, V=0.5(V), V=1.5(V)  in sequence.
  }
\label{Fig.2}
\end{figure}

  It seems also possible to set up an experiment with negative
impurity charge and negative  tip voltage close to the value $\varepsilon_d
+ U$. In this case $W > 0$, and the Coulomb corrections
to the tunnelling amplitude
result in power-low behavior of the tunnelling current with the opposite sign
exponent in Eq.(\ref{If}). The tunnelling current is suppressed near the threshold, compared
to the noninteracting case. This behavior is shown in Fig. 2b.

 Experimental
STM/STS investigations of deep impurity levels on semiconductor surfaces
give evidence of the existence of the described effects.  Some STM images
demonstrate non-monotonic dependence of tunnelling current on
applied bias voltage
~\cite{1}.  With increasing of bias voltage impurity atom is "switched on" in
STM image - appears as a bright spot. But further increase of the
tunnelling bias "extinguishes" the brightness of the impurity atom and
it is seen as a dark spot in the STM image (see inset in Fig.2).
According to the present model it can be explained by
decrease of the tunnelling current caused by the Coulomb vertex
corrections to the tunnelling amplitude for some types of impurities.

We thank L.V.Keldysh for helpful discussions.
This work was supported by the RBRF grants N 00-15-96558, 00-02-17759
and program "Nanostructure"  .

\end{document}